\documentstyle[aps,pre,epsfig,multicol,amssymb]{revtex}
\begin{document}
\draft

 \title{Effective forces in  colloidal mixtures: from depletion
 attraction to accumulation repulsion} 
\author{A.A. Louis$^{a}$, E. Allahyarov$^b$, H. L\"{o}wen$^b$,
 and  R.Roth$^{c}$ }
\address{$^a$Department of Chemistry, Lensfield Rd, Cambridge CB2 1EW,
UK}
 \address{$^b$Institut f\"{u}r Theoretische Physik
II, Heinrich-Heine-Universit\"{a}t D\"{u}sseldorf, D-40225
D\"{u}sseldorf, Germany} 
\address{$^c$
Max-Planck Institut f{\"u}r 
Metallforschung, Heisenbergstrasse 1, D-70569 Stuttgart, Germany and ITAP,
University of Stuttgart, Pfaffenwaldring 57, D-70569 Stuttgart, Germany
}
 \date{\today} \maketitle
\begin{abstract}
\noindent
 Computer simulations and theory are used to systematically
investigate how the effective force between two big colloidal spheres
in a sea of small spheres depends on the basic (big-small and
small-small) interactions.  The latter are modeled as hard-core pair
potentials with a Yukawa tail which can be both repulsive or
attractive.  For a repulsive small-small interaction, the effective
force follows the trends as predicted by a mapping onto an effective
non-additive hard-core mixture: both a depletion attraction and an
accumulation repulsion caused by small spheres adsorbing onto the big
ones can be obtained depending on the sign of the big-small
interaction.  For repulsive big-small interactions, the effect of
adding a small-small attraction also follows the trends predicted by
the mapping.  But a more subtle ``repulsion through attraction''
effect arises when both big-small and small-small attractions occur:
upon increasing the strength of the small-small interaction, the
effective potential becomes more repulsive.  We have further tested
several theoretical methods against our computer simulations: The
superposition approximation works best for an added big-small
repulsion, and breaks down for a strong big-small attraction, while
 density functional theory is very accurate for any big-small
interaction when the small particles are pure hard-spheres.  The
theoretical methods perform most poorly for small-small attractions.
\end{abstract} 

\begin{multicols}{2}
\section{Introduction}

A fundamental description of colloidal interactions based on
statistical mechanics is needed to understand and predict the
stability and the phase behavior of colloidal
suspensions\cite{Russ89,PuseyLH}.  Most
inter-colloidal forces are effective in the sense that some
microscopic degrees of freedom are averaged out.  This concept of mean
or effective interactions \cite{Bell00,Loui01a,Liko01} is crucial to
bridge the different length scales involved in colloidal systems and
has been exploited in many different circumstances.  Examples of the
microscopic degrees of freedom include solvent
particles\cite{solvent}, smaller colloidal particles\cite{Croc99},
added polymer coils \cite{Asak54,Asak58,Lekk92,Ilet95} or monomers of
(grafted) polymer chains\cite{Liko98,Loui00b,Bolh01a,Bolh01b} as well
as counter-ions\cite{Lowe93,Alla98} and salt ions\cite{Trig98} in the
case of charged suspensions.  The resulting effective forces turn out
to exhibit a wide range of features: They can be attractive,
repulsive, or oscillatory, and are an important key to understanding
colloidal stability as well as flocculation and coagulation
\cite{Russ89}.

\begin{figure}
\begin{center}
\epsfig{figure=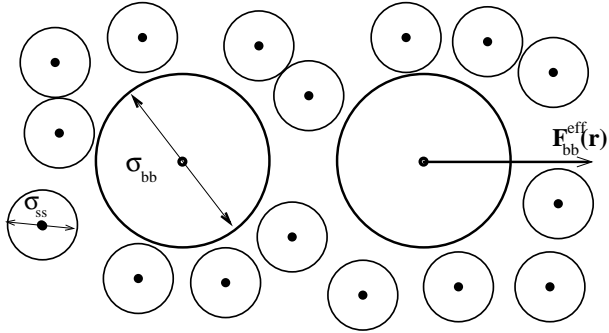,angle=-90,width=8cm} 
\caption{\label{fig:schematic} Two big spheres of radius $\sigma_{bb}$
experience an effective force $F_{bb}^{eff}(r)$ induced by the sea of
small spheres or radius $\sigma_{ss}$ .
}
\end{center}
\end{figure}

In the present paper we consider a binary colloidal mixture of big and
small colloidal particles and investigate the distance-resolved
effective force and potential between two big colloidal spheres
surrounded by many small colloidal spheres, as depicted in
Fig.~\ref{fig:schematic}. The small particles are averaged out so that
one is left with an effective pair potential between the big ones,
which is superimposed onto the direct big-big interaction.  The form
of these effective interactions depends sensitively on the basic
big-small and small-small interactions.

 Most previous studies have focused on a hard-sphere (HS)-like
interaction between the colloidal particles, with a special emphasis
on two cases: {\bf (1)} the Asakura-Oosawa model, which was originally
designed to describe mixtures of sterically-stabilized colloids and
nonadsorbing polymer coils \cite{Asak58}.  In this strongly
non-additive\cite{nonadd} model, the big-small interaction is HS like,
while the small-small interaction is ideal gas like. The effective
pair potential between two colloids can be exactly calculated and has
an attractive well proportional to the density of the small particles,
and a range equal to their diameter.  {\bf (2)} Much work has also
focused on the complementary model of additive HS mixtures, inspired
in part by a prediction that they might phase-separate\cite{Bibe91}.
The effective depletion forces were obtained by computer simulation
\cite{Bibe96,Dick97}, approximate theories\cite{Walz94,Mao95}, density
functional theory (DFT) \cite{Gotz99,Roth00a}, and
experiments\cite{Croc99,Rudh98,Bech99}.  Simplified potentials were
used to investigate the phase-behavior of binary HS
mixtures\cite{Dijk98}, where it was found that fluid-fluid
phase-separation, when it occurs\cite{Loui01b,Roth01b}, was always
metastable w.r.t.\ fluid-solid phase-separation.

  In this work we generalize these studies to arbitrary big-small and
small-small interactions.  Our study is motivated by the fact that
real colloidal suspensions typically have soft interactions which are
beyond the HS model. These soft interactions can easily be
tuned, for example by changing the solvent composition, the amount of added
salt, the surface charge, etc..., which provides a route to tailoring
the effective interactions between the big particles.  Hence it is
essential to understand on a full statistical level how the basic
big-small and small-small interactions affect the effective big-big
interactions in a binary colloidal mixture.

 There has been some previous work on how non-HS like interactions
affect the effective big-big interactions, see e.g.\
refs.~\cite{Bell00,Mond95,Gari97,Amok98,Malh99,vonG99,Clem00,Mend00,Piec00,Kuwa00,Egor01,Amok01,Gonz01},
which showed that the differences with the pure HS case could be
substantial.  However, with a few exceptions, all these studies were
performed with approximate methods such as integral equations.  Our
goal here is to provide benchmark {\it exact} computer simulation
results for a set of $9$ different parameters corresponding to
combinations of attractive and repulsive small-small and big-small
interactions. This allows us to {\em systematically} study the trends
in the effective forces as induced by the basic interactions.  It also
allows us to test several theoretical techniques, namely DFT, the
superposition approximation\cite{Atta89}, and a mapping to a
non-additive HS model\cite{Loui01b}.

Generalizing from an additive to a non-additive HS mixture already
leads to a much richer class of effective
potentials\cite{Loui00a,Roth01a,Loui01b,Roth01b}, which can be exactly
calculated from an accurate theory of additive HS
mixtures\cite{Roth00a}.  Similarly, we find here that the effective
big-big interactions are profoundly influenced by the basic big-small
and small-small interactions.  The trends can be summarized as
follows: Adding a big-small repulsion results in enhanced attraction
through the standard depletion mechanism\cite{Asak54,Asak58}. Adding a
big-small attraction leads to an accumulation of the small particles
near each big one, which in turn results in a more repulsive big-big
interaction (accumulation repulsion).  For a given big-small
interaction, adding a small-small repulsion also results in an
enhanced small particle density near the big particles, and therefore
in a more repulsive big-big interaction.  Adding a small-small
attraction when the big-small interaction is repulsive results in a
more attractive effective interaction.  All these effects can be
qualitatively understood from a mapping to a non-additive HS Model.
But when a small-small and a big-small attraction are combined, a {\em
repulsion through attraction} effect occurs, which is not captured by
DFT or by our mapping scheme.

The paper is organized as follows: In Chapter II, we discuss how to
formally map a two-component mixture onto an effective one-component
one, and define our target quantities. In Chapter III, we describe our
model for the basic interactions.  We discuss our simulation results
in Chapter IV, and the results of several different theories in
Chapter V.  Our conclusions are stated in Chapter VI.

\section{Mapping a binary mixture onto an effective one-component system}
\label{sec:mapping}

In this section we briefly describe how to map a two-component binary
mixture onto an effective one-component system.  This statistical
mechanical procedure was originally developed by McMillan and
Mayer\cite{McMi45} to trace out the solvent in a suspension, but it
can be used just as well to trace out a smaller component in a binary
colloid mixture.

We begin with the Hamiltonian of our two-component mixture
\begin{equation}
H = K + H_{bb} + H_{ss} + H_{bs}.
\end{equation}
Here $K$ is the total kinetic energy of the mixture and three
 potential energy contributions are:
\begin{eqnarray}\label{eqa.2}
H_{bb} &=& \sum_{i<j}^{N_b} \Phi_{bb}({\bf r}_i^b-{\bf
r}_j^b),\nonumber \\ H_{ss} &=& \sum_{i<j}^{N_s} \Phi_{ss}({\bf
r}_i^s-{\bf r}_j^s), \nonumber \\ H_{bs} &=& \sum_{i=1}^{N_b}
\sum_{j=1}^{N_s} \Phi_{bs}({\bf r}_i^b-{\bf r}_j^s),
\end{eqnarray}
where the $\Phi_{ij}$ denote the pairwise interaction potentials, and
${\bf r}_i^b$ and ${\bf r}_j^s$ denote the coordinates of the centers
of the  big and small particles respectively.

The formal procedure to reduce the two-component system to an
effective one-component one, described in more detail elsewhere, see
e.g.\cite{Loui01a,Liko01,Lowe93,Dijk98,Roth01b}, consists of tracing out the
small particles for a fixed configuration $\{{\bf r}_i^b\}$ of the big
particles.  This is most conveniently done for a semi-grand ensemble
where the small particles in the volume $V$ are kept at a fixed
chemical potential $\mu_s$.  The resulting effective one-component
Hamiltonian that governs the behavior of the big particles takes the
form:
\begin{equation}\label{eqa.4}
H_{bb}^{eff} =H_{bb} + \Omega,
\end{equation}
where $\Omega=\Omega(N_b,z_s,V;\{{\bf r}_i^b\})$ is the grand
potential of a fluid of small particles at fugacity $z_s =
\lambda_s^{-3}\exp(\beta \mu_s)$, ($\beta =1/k_B T$) subjected to the
external potential of the fixed configuration $\{{\bf r}_i^b\}$ of the
big particles. This grand-potential can be further expanded as a sum
of (n-body) terms\cite{Dijk98}:
\begin{equation}\label{eqa.5}
\Omega = \sum_{n=0}^{N_b} \Omega_n.
\end{equation}
The first term, $\Omega_0$, is the so-called volume term, which only
depends on properties of the small-particles (i.e.\ $\Phi_{ss}$) and
is therefore independent of the configuration $\{{\bf r}_i^b\}$. The
one-body term $\Omega_1$ can be related to $N_b$ times the free-energy
gained by inserting a single big particle into a sea of small
particles.  It depends on both $\Phi_{ss}$ and $\Phi_{bs}$, but is
independent of $\Phi_{bb}$. Of most interest to us here is the
two-body term $\Omega_2$, which can be written as:
\begin{equation}\label{eqa.6}
\Omega_2(N_b,z_s,\{{\bf r}_i^b\}) = \sum_{i<j}^{N_b}
V_{bb}^{eff}(|{\bf r}_i - {\bf r}_j|),
\end{equation}
where the two-body effective pair-potential $V_{bb}^{eff}(r)$ is
related to the free-energy difference between two big spheres a
distance $r$ apart, immersed in a sea of small spheres at fugacity
$z_s$, and the same system with the two spheres at $r=\infty$.  It is
important to notice that $V_{bb}^{eff}(r)$ itself only depends on
$\Phi_{bs}$ and $\Phi_{ss}$, and not directly on $\Phi_{bb}$.  This
observation was a key to understanding why the effective potentials
for non-additive HS mixtures can be calculated by a theory for
additive HS mixtures\cite{Loui01b,Roth01b,Roth01a}. Nevertheless, through
Eq.~(\ref{eqa.4}), the effect of $\Phi_{bb}(r)$ is felt in that it sets
the range where $V^{eff}_{bb}(r)$ is relevant.

In a similar way, higher order interactions can be
derived\cite{Bolh01b}, but for this scheme to remain tractable, one
usually truncates at the pair level.  As long as the ranges of
$\Phi_{ss}$ and $\Phi_{bs}$ are small compared to the range of
$\Phi_{bb}$, higher order terms are not expected to be very
important\cite{Dijk98,Roth01b}.

Thus far we have developed this formal tracing out in the semi-grand
ensemble, which helps emphasize that in a mixture with many large
particles, the correct effective potential $V^{eff}_{bb}(r)$ to be
used in Eq.~(\ref{eqa.6}) is the one fixed by the chemical potential
or fugacity instead of the overall density of the smaller particles,
as first emphasized by Lekkerkerker {\em et al.}\cite{Lekk92}.  It is
important to keep this in mind when studying phase-behavior.
However, for a practical calculation of the pair term, with only two
big particles in an infinite sea of small ones, one can just as easily
work in the canonical ensemble, keeping in mind that when using the
obtained $V_{bb}^{eff}(r)$ for a system with many big particles, the
input density $\rho_s$ is really that of a reservoir of small
particles kept at the same chemical potential as the full mixture of
big and small particles\cite{Lekk92,Dijk98,Roth01b}.

For the case of two big-spheres a distance $r$ apart, in a bath of
small particles at density $\rho_s$, the average force is related to
the effective potential by:
\begin{equation}\label{eqa.7}
{\bf F}_{bb}^{eff}(r) = - \frac{\partial}{\partial {\bf r}}
V_{bb}^{eff}(r).
\end{equation}
This average force provides an intuitive way of understanding the
effective interactions which parallels the more formal derivation
above.  First we define a one-body density as follows:
\begin{equation}\label{eqa.8}
\rho^{(1)}_s({\bf r}') = << \sum_i^{N_s} \delta({\bf r}'-{\bf r^s_i})>>,
\end{equation}
where $<<>>$ denotes a canonical thermodynamic average over the small
particles.  If one big particle is fixed at the origin, and one is
fixed at {\bf r}, then we can define a one-particle density of the
small particles $\rho_s^{(1)}({\bf r}';{\bf 0}, {\bf r})$ for that
fixed configuration.  The effective force induced between two big
spheres by the smaller particles can then be written
as\cite{Atta89,Bibe96}:
\begin{equation}\label{eq3.1}
{\bf F}_{bb}^{eff}({\bf r}) = -\int \rho_s^{(1)}({\bf r}';{\bf 0}, {\bf
r}) \frac{\partial}{\partial {\bf r}'} \phi_{bs}(r') d{\bf r}',
\end{equation}
which has the obvious physical interpretation that the total effective
or mean force is simply the average of the sum over all the big-small
interactions.

In the next sections, we will concentrate on the effect of changing
$\Phi_{ss}(r)$ and $\Phi_{bs}(r)$, at fixed reservoir density
$\rho_s$, on the effective interactions $F^{eff}_{bb}(r)$ and
$V^{eff}_{bb}(r)$.

\section{Models}

\subsection{Interactions}

We model the basic small-small and big-small interactions defined in
Eq.~(\ref{eqa.2}) as hard-core Yukawa pair potentials:
\begin{equation} \label{eqb.3}
\Phi_{ij}(r) = \left\{
\begin{array}{lr}
\infty & r< \sigma_{ij}\\
\phi_{ij}(r) & \mbox{otherwise},
\end{array}
\right.
\end{equation}
where in each case $r$ denotes the distance between the centers of the
relevant particles. Throughout, we take $\phi_{bb}(r)=0$, while the
big-small interaction added to the HS repulsion is given by:
\begin{equation}\label{eq2.1}
\phi_{bs}(r) = \frac{ \epsilon_{bs} \sigma_{bs}}{r} \exp \left[ -
\kappa_{bs} (r - \sigma_{bs}) \right],
\end{equation}
and the small-small interaction added to the HS repulsion is given by:
\begin{equation}\label{eq2.2}
 \phi_{ss}(r) = \frac{ \epsilon_{ss} \sigma_{ss}}{r} \exp \left[ -
\kappa_{ss} (r - \sigma_{ss}) \right],
\end{equation}
where $\sigma_{bs} = \frac12 (\sigma_{bb} + \sigma_{ss})$.
%

 The Yukawa tail can be either repulsive or attractive. This gives us
the possibility to change the interaction over a broad range from pure
HS (which serve as a reference case) to soft repulsions and attractive
tails. We note that a hard-core Yukawa interaction has been frequently
used in theoretical studies on liquids, see e.g.~\cite{Cacc96} and
refs therein. Both an attractive and repulsive tail is realized in
various colloidal solutions: a short-ranged attractive Yukawa tail has
been shown to satisfactorily model the ``stickiness" in the
interaction between globular protein solutions \cite{Cost00}. A
repulsive Yukawa tail, on the other hand, describes charged
suspensions \cite{PuseyLH,Kram93} where the range is controlled by the
added salt concentration and the amplitude by the colloidal charge.

\subsection{Parameter combinations}

 Our aim is to systematically investigate the effect of the basic
big-small and small-small interactions on the effective big-big
interactions.  We vary the big-small interaction by changing $\beta
\epsilon_{bs}$, and vary the small-small interaction by changing
$\beta \epsilon_{ss}$. The other parameters are fixed as follows:
$\sigma_{ss} = 0.2 \sigma_{bb} $, $\kappa_{bs} = 6/\sigma_{bb} =
1.2/\sigma_{ss}$, and $\kappa_{ss} = 15/\sigma_{bb} = 3/\sigma_{ss}$.
The range of the big-small interaction is of the order of the small
particle size, while the range of the small-small interactions is
significantly less than the small particle size. For all simulations
the packing fraction of the small particles in the bulk is set to
$\eta_s = \pi \rho_s \sigma_{ss}^3/6 = 0.1$.

In total, we studied $9$ different parameter combinations
corresponding to taking $\beta \epsilon_{bs}$ and $\beta
\epsilon_{ss}$ to be positive, negative, or zero.  The detailed values
are given in table~\ref{tableI}.

\section{Computer simulations of effective forces and interactions}\label{sec:simulation}
\subsection{Simulation method}

Our simulation set-up contains two big spherical particles in a large
cubic simulation box of length $L$ with periodic boundary conditions
in all three directions.  The big particles with their
center-to-center separation $r$ are fixed along the body diagonal of
the cubic box and $N_s=10000$ small mobile particles of diameter
$\sigma_{ss}$ are added to the box. The box length $L=7.49
\sigma_{ss}$ is sufficiently large to exclude any spurious periodic
image effects.  We also studied a single big particle in the
simulation box to access the one-body small-particle density
$\rho_s(r)$ around an isolated big sphere. For two big spheres $\rho_s
= N_s/(L^3 - \frac13 \pi \sigma_{bb}^3)$, while for one big sphere
$\rho_s = N_s/(L^3 - \frac16 \pi \sigma_{bb}^3)$. We checked the
asymptotic density profiles to confirm that the bulk densities were
always the same.

We use a molecular dynamics (MD) code combining the velocity Verlet
algorithm\cite{Alle91} with discrete collisions and reflections
induced by the hard core of the interaction potentials in order to
calculate the trajectories of the small particles. The system was
carefully equilibrated and then statistical averages were computed
such as the mean force acting on the big particles (see
Eq.~(\ref{eq3.1})) or the inhomogeneous density field $\rho_s(r)$ of the
small particles around a single big particle.  Details of our
simulation procedure are described in references\cite{Trig98,Alla01}.

\subsection{One body density profiles}

Before we discuss the depletion potentials it is useful to first
examine the density profiles $\rho_s(r)$ of the small particles around
a single sphere.

\subsubsection{No extra big-small interaction: runs 1-3}

\begin{figure}
\begin{center}
\epsfig{figure=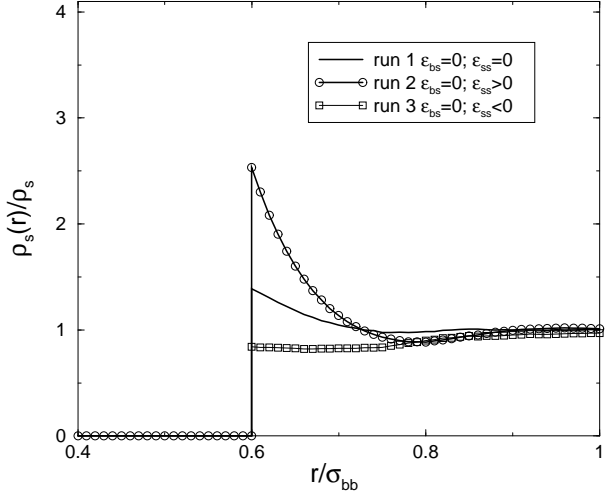,width=8cm} 
\caption{\label{fig:hrB0-new}Normalized density profiles $\rho_s(r)/\rho_s$
of centers of the small spheres as a function of the distance $r$ from
the center of a single big sphere.  Results are from computer
simulations for runs $1-3$; i.e.\ $\epsilon_{bs}=0$, and
$\epsilon_{ss}$ is varied. The values of the $\epsilon_{ij}$ can be
found in table~\protect\ref{tableI}.  In this figure, as well as
Figs.~3-6, the circles denote a repulsive $\epsilon_{ss}$, the squares
an attractive $\epsilon_{ss}$ and no symbol means no added
$\epsilon_{ss}$. Solid lines denote no added $\epsilon_{bs}$, dotted
lines $\epsilon_{bs}>0$, and dot-dashed lines $\epsilon_{bs}<0$.  }
\end{center}
\end{figure}

In Fig.~\ref{fig:hrB0-new}, these are shown for runs $1-3$, where
$\epsilon_{bs}=0$.  Compared to the density profile of the pure HS,
there is a significantly increased accumulation near the big spheres
when a small-small repulsive interaction is added.  To first order
this can be understood by mapping the $\phi_{ss}(r)$ onto an effective
HS diameter $\sigma_{ss}^{eff}$\cite{Hans86}: Adding a small-small
repulsive interaction increases the effective HS size, and therefore
also the effective HS packing fraction, resulting in a more pronounced
accumulation of density near the surface.  Adding a small-small
attraction has the opposite effect.  This is because the bulk is now
more favorable for the small particles, and they are attracted to it.

\subsubsection{Added big-small repulsion: runs 4-6}

In Fig.~\ref{fig:hrBrepulsive-new}, the density profiles are shown for
runs $4-6$, where $\beta \epsilon_{bs}=0.82$. As expected, the
repulsive big-small interaction leads to a reduction of the density at
contact.  Again, adding a small-small repulsion increases the density
at contact, and adding a small-small attraction leads to a further
depletion of the density at contact.

\subsubsection{Added big-small attraction: runs 7-9}

\begin{figure}
\begin{center}
\epsfig{figure=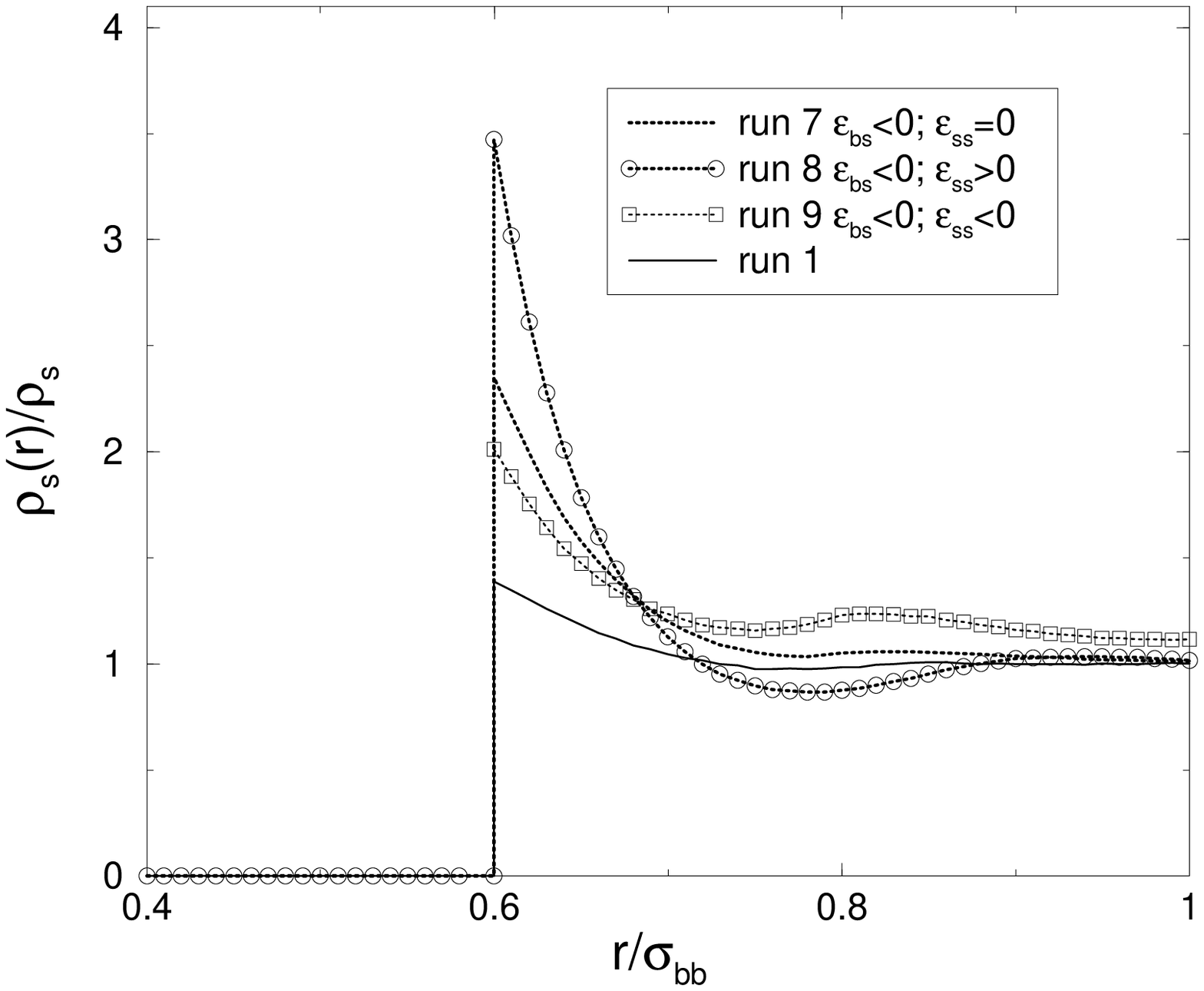,width=8cm} 
\caption{\label{fig:hrBattractive-new} Normalized density profiles
$\rho_s(r)/\rho_s$ from computer simulations for runs $7-9$;
$\epsilon_{bs} < 0$, $\epsilon_{ss}$ is varied.  We also show the
result for run 1, the pure HS case. }
\end{center}
\end{figure}

In Fig.~\ref{fig:hrBattractive-new}, the density profiles are compared
for runs $7-9$, where $\beta \epsilon_{bs} = -0.82$.  As expected, the
attractive big-small interaction results in an enhanced density at
contact.  Similarly to the previous two cases where $\epsilon_{bs} =0$
and $\epsilon_{bs} > 0$, respectively, adding a small-small repulsion
enhances the contact density w.r.t.\ the case of no small-small
repulsion.  But in contrast to the two previous cases, where adding a
small-small attraction resulted in a depleted density profile w.r.t.\
the pure HS case, here adding $\epsilon_{ss}<0$ results in an
enhanced total accumulation of the small particles near the big one.
Although the contact value is slightly smaller than the case for pure
HS, there is a marked accumulation further out, corresponding to a
second layer of particles.  The relative adsorption of the small
particles around the big one is thus considerably larger than for the
case of no small-small attractions.  This can be understood by the
following simple argument: the big-small attraction leads to an
accumulation of the small particles near the surface of the big
sphere. When this accumulation is large enough, it becomes favorable
for the small particles to leave the bulk, and approach the surface of
the big particle, where their local density is larger.  Thus the two
attractions amplify each other in a non-linear fashion.

In conclusion then, when comparing Figs.~\ref{fig:hrB0-new},
\ref{fig:hrBrepulsive-new} and \ref{fig:hrBattractive-new}, it is clear that
for the parameters we have chosen, the big-small interaction has the
largest relative effect on the density profiles.  The effect of adding
a small-small repulsion can be qualitatively understood by the larger
effective sphere size and concomitant larger packing
fraction.  The effect of adding a small-small attraction can be
qualitatively understood by the fact that the bulk is usually
preferred over the surface of the particle, except when the big-small
interaction is strong enough to provoke a non-linear enhancement of
the density of small particles near a big one.

Having investigated the effect of the interactions on the one-body
density profiles, we now turn to the related two-body depletion
potentials.

\subsection{Effective pair forces and potentials}

\begin{figure}
\begin{center}
\epsfig{figure=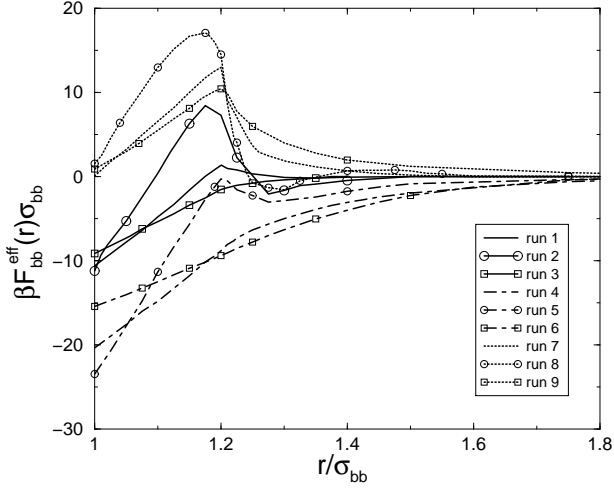,width=8cm} 
\caption{\label{fig:allforces-new} Effective forces between the big
particles, $\beta F_{bb}^{eff}(r)$, shown for all 9 runs.  The symbols are the
same as in Figs.~2-4. The results are from direct computer simulations.
Notice that the big-small interaction has the dominant effect. A
repulsive $\epsilon_{bs}$ results in a more repulsive $\beta F_{bb}^{eff}(r)$
(dot-dashed lines), while an attractive $\epsilon_{bs}$ results in a
more attractive $\beta F_{bb}^{eff}(r)$ (dotted lines).  The small-small
interactions have a relatively smaller effect.  }
\end{center}
\end{figure}

\begin{figure}
\begin{center}
\epsfig{figure=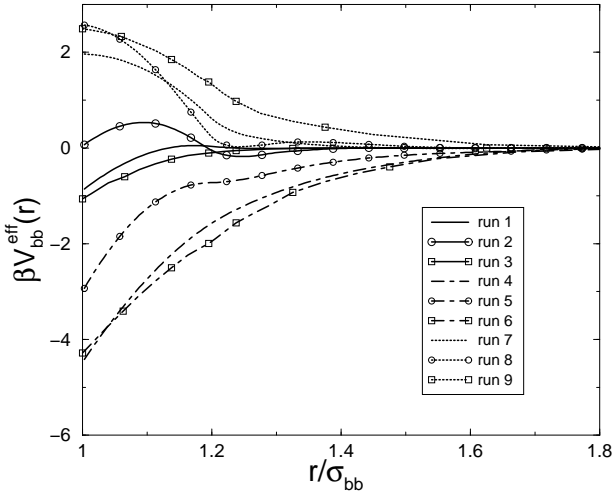,width=8cm} 
\caption{\label{fig:Velshad-new} Effective potentials between the big
particles, $\beta V_{bb}^{eff}(r)$, shown for all 9 runs.  The results are from
direct computer simulations. The symbols are the same as in Figs.~2-5.  }
\end{center}
\end{figure}

In Fig.~\ref{fig:allforces-new} we compare the effective depletion force
between two big spheres for all nine parameter combinations detailed
in table~\ref{tableI}.  In Fig.~\ref{fig:Velshad-new} we compare the
related depletion potentials.  First we note that changing the
big-small interaction is the dominant effect: the depletion forces and
potentials split naturally into three groups: no added big-small
interaction (solid lines, runs $1-3$), big-small attraction (dotted
lines, runs $4-6$), and big-small repulsion (dot-dashed lines, runs
$7-9$).  We will treat each case in turn.

\subsubsection{No extra big-small interaction: runs 1-3}

As was already seen for the one-body profiles, adding a small-small
repulsion (run 2) results in a larger effective small-sphere size and
packing fraction, which is reflected in more pronounced oscillations
compared to the pure HS case (run 1). These are evident both in the
effective force and in the effective pair potential.  Adding a
small-small attraction (run 3) results in a reduced density near a big
particle, as seen in Fig.~\ref{fig:hrB0-new}. This would imply that
each big sphere excludes slightly more free volume than if there were
no small-small attractions, which, in turn, implies a slightly more
attractive potential, as is observed.

\subsubsection{Added big-small repulsion: runs 4-6}

The dominant effect of adding a big-small repulsion is to make the
effective forces and potentials much more attractive.  This can be
understood with the classical picture of
depletion\cite{Asak54,Asak58}: Adding a big-small repulsion results in
a larger depletion layer around each large particle.  When two large
spheres approach, the amount of doubly excluded volume is therefore
larger, resulting in a more attractive effective (depletion) potential
or force.  Again, adding a small-small repulsion (run 5) results in
enhanced layering as compared to the pure HS small particles.  Adding
a small-small attraction (run 6) has only a weak effect similar to
what was seen for run 3.

\subsubsection{Added big-small attraction: runs 7-9}

The dominant effect of adding a big-small attraction is to make the
effective forces and potentials much more repulsive.  This can be
qualitatively understood from the fact that the big-small attraction
results in an enhanced density of small particles near a single large
sphere.  When two such large spheres approach one another, the layers
of small particles around each one begin to interact, leading to an
enhanced repulsion between them.  As demonstrated in
Fig.~\ref{fig:hrBattractive-new} for a given big-small attraction,
adding small-small repulsion or attraction both result in a further
enhanced density $\rho_s(r)$ of the small particles around a single
large one, which in turn explains why the effective pair potentials
are more repulsive for both $\epsilon_{ss} >0$ and $\epsilon_{ss} <0$.
Relatively speaking, adding small-small attractions has the largest
effect on the effective potentials, which is due to the non-linear
coupling between the big-small and the small-small attractions.

\section{Theoretical descriptions of the effective forces and potentials}

\subsection{Superposition approximation}

\begin{figure}
\begin{center}
\epsfig{figure=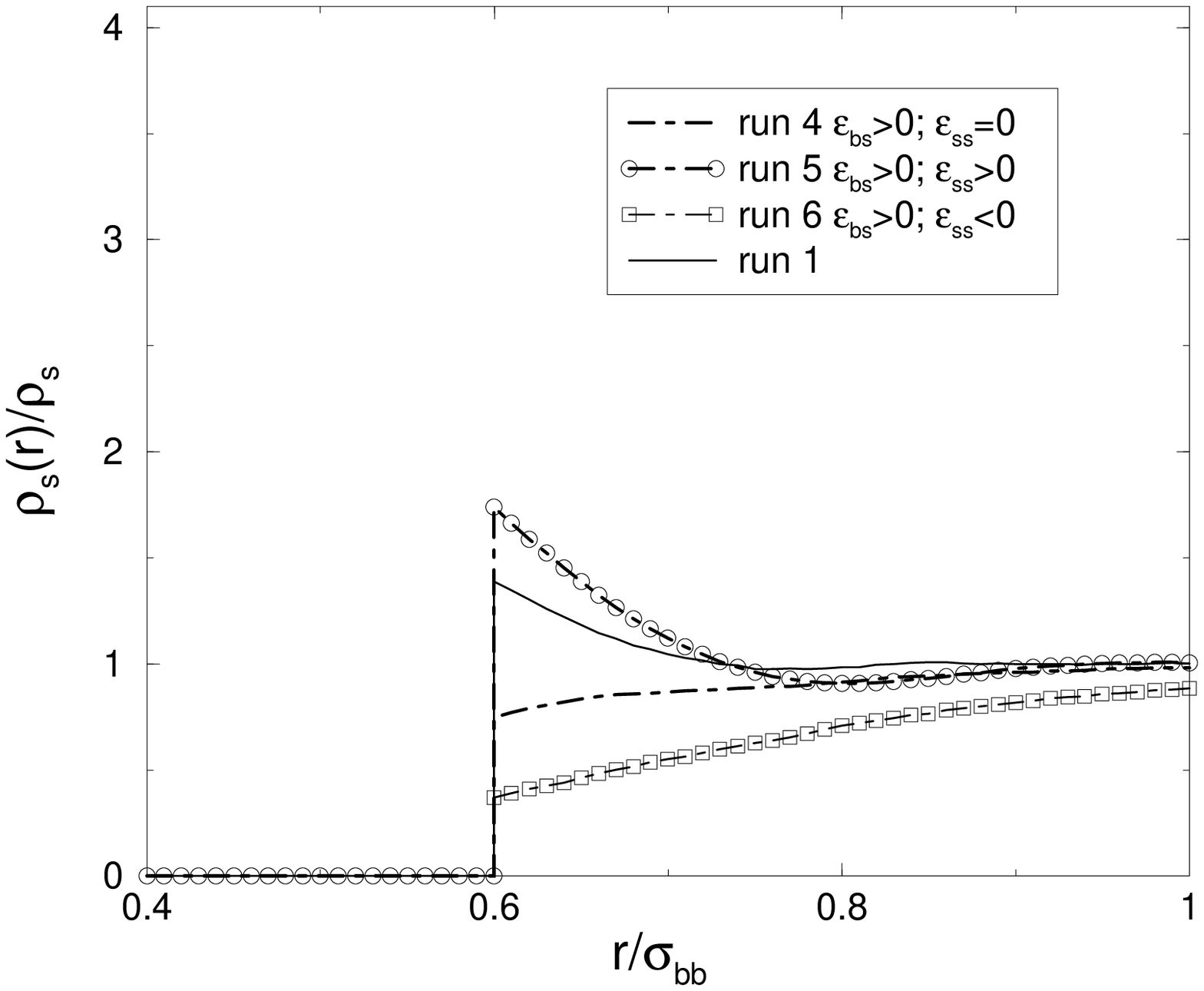,width=8cm} 
\caption{\label{fig:hrBrepulsive-new} Normalized density profiles
$\rho_s(r)/\rho_s$ from computer simulations for runs $4-6$;
$\epsilon_{bs} > 0$, $\epsilon_{ss}$ is varied. We also show the
result for run 1, the pure HS case.  }
\end{center}
\end{figure}

\begin{figure}
\begin{center}
\epsfig{figure=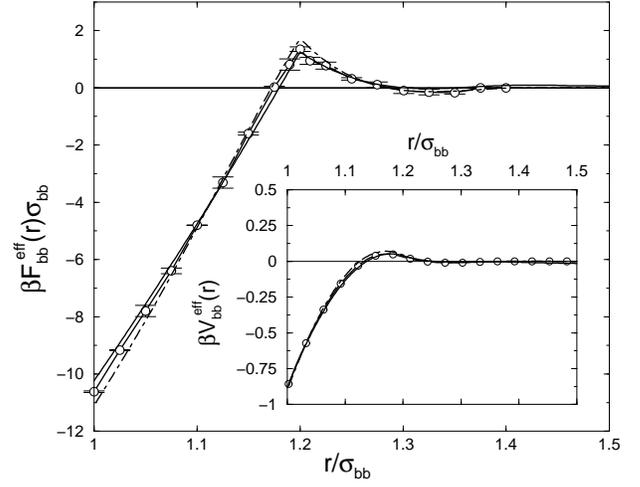,width=8cm} 
\caption{\label{fig:force1} Comparison of theory to simulation for run
1: $\epsilon_{bs}=0$, $\epsilon_{ss}=0$. Simulation (solid line
with symbols), superposition approximation (solid line), and DFT
(dot-dashed line) results for the effective force and potential (see
inset) as a function of $r$, the distance between the centers of the
big particles, are shown.}
\end{center}
\end{figure}

\begin{figure}
\begin{center}
\epsfig{figure=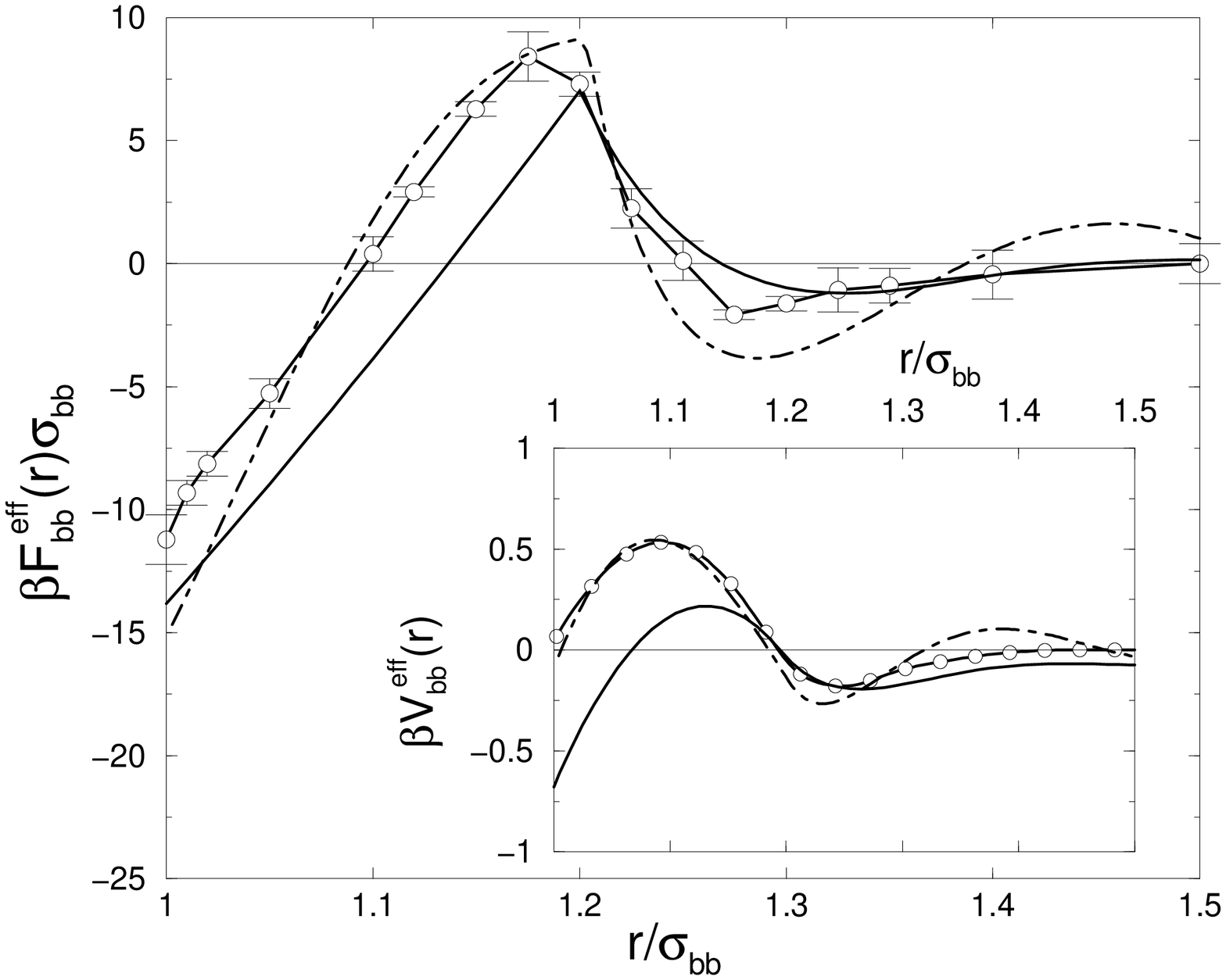,width=8cm} 
\caption{\label{fig:force2}  Same as Fig.~\ref{fig:force1}, but for
run2:
$\epsilon_{bs}=0$, $\epsilon_{ss}>0$.
}
\end{center}
\end{figure}

\begin{figure}
\begin{center}
\epsfig{figure=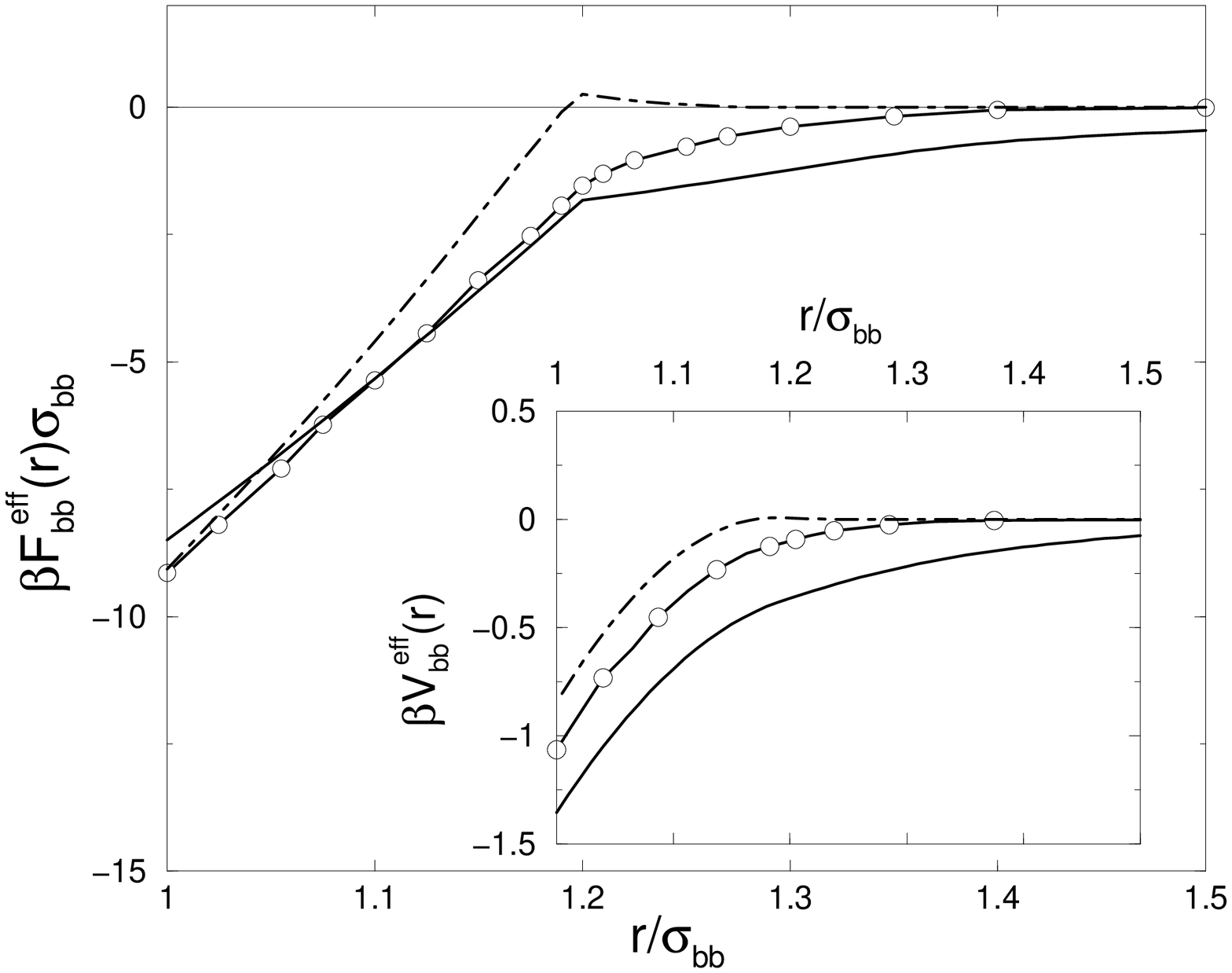,width=8cm} 
\caption{\label{fig:force3} Same as Fig.~\ref{fig:force1}, but for run3: 
$\epsilon_{bs}=0$, $\epsilon_{ss}<0$.
}
\end{center}
\end{figure}

\begin{figure}
\begin{center}
\epsfig{figure=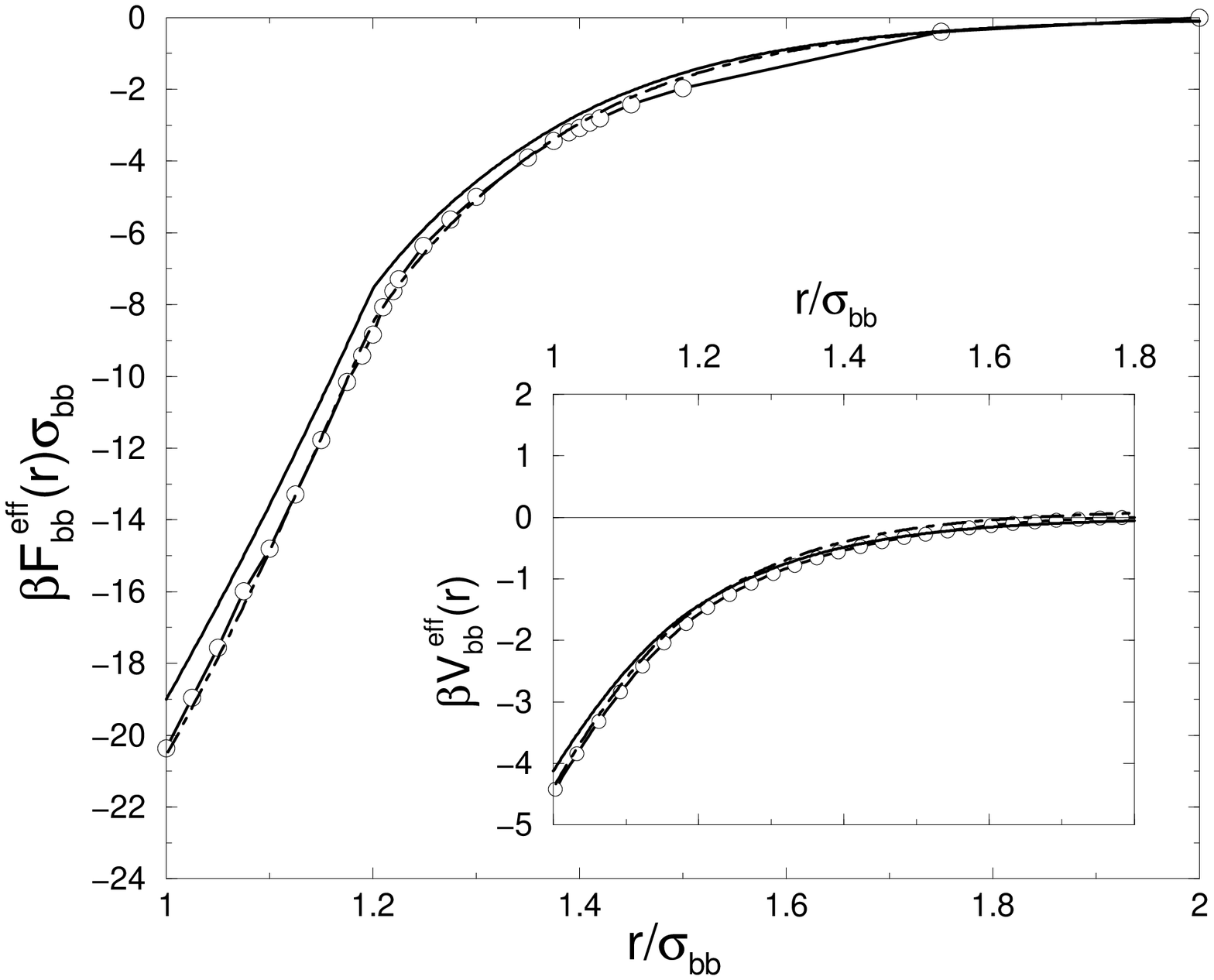,width=8cm} 
\caption{\label{fig:force4} Same as Fig.~\ref{fig:force1}, but for run4: 
$\epsilon_{bs}>0$, $\epsilon_{ss}=0$.
}
\end{center}
\end{figure}

\begin{figure}
\begin{center}
\epsfig{figure=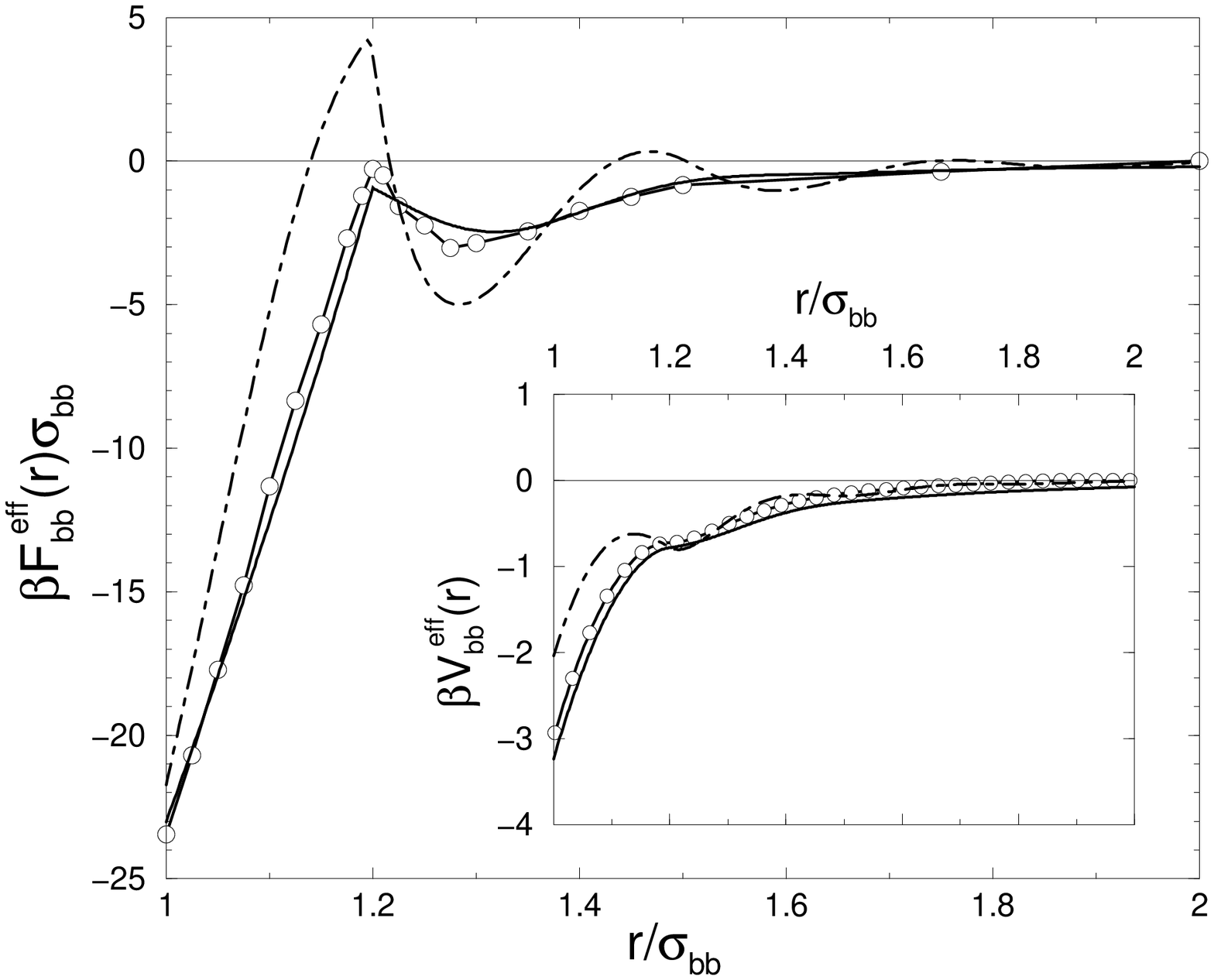,width=8cm} 
\caption{\label{fig:force5} Same as Fig.~\ref{fig:force1}, but for run5: 
$\epsilon_{bs}>0$, $\epsilon_{ss}>0$.
}
\end{center}
\end{figure}

\begin{figure}
\begin{center}
\epsfig{figure=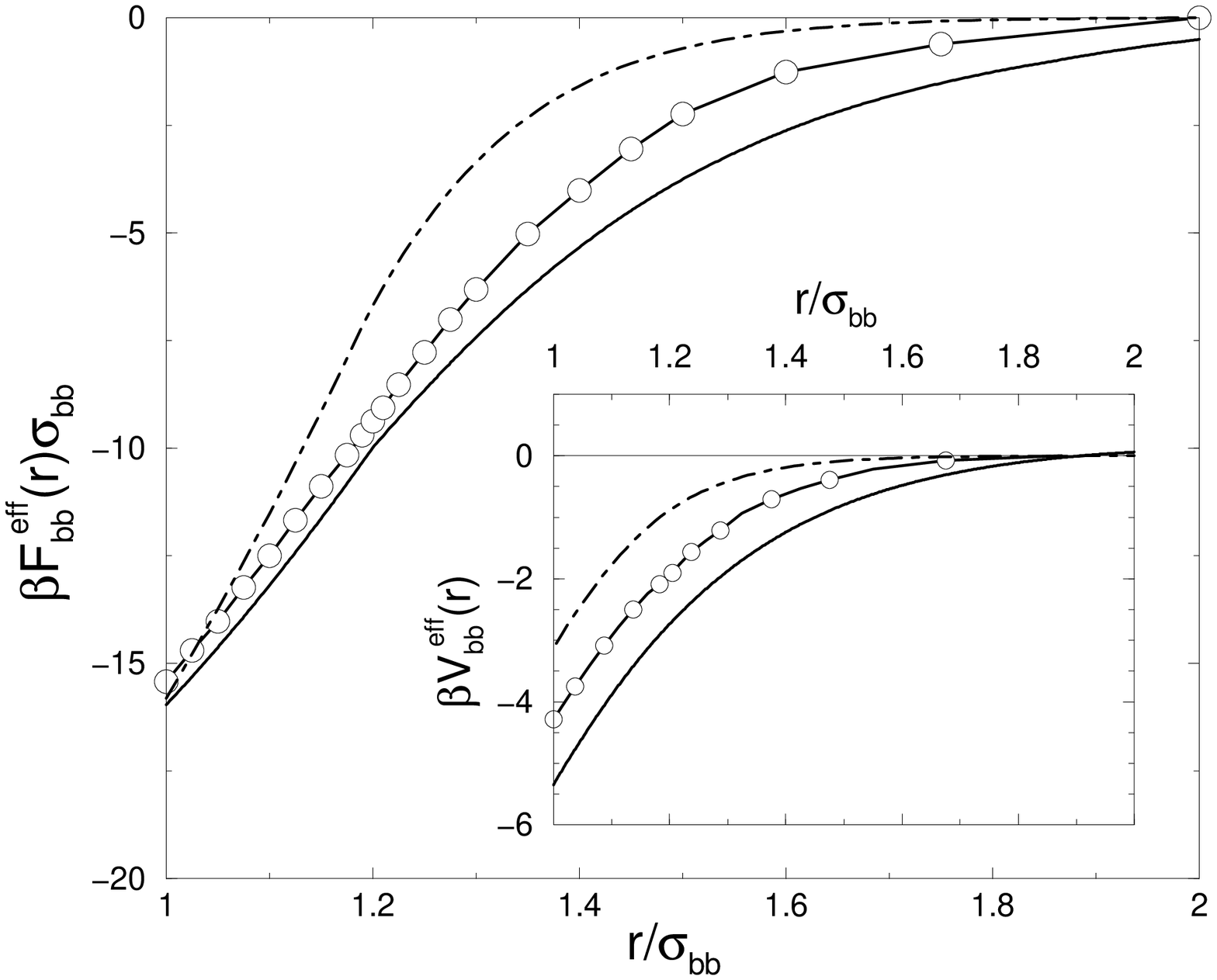,width=8cm} 
\caption{\label{fig:force6} Same as Fig.~\ref{fig:force1}, but for run6: 
$\epsilon_{bs}>0$, $\epsilon_{ss}<0$.
}
\end{center}
\end{figure}

\begin{figure}
\begin{center}
\epsfig{figure=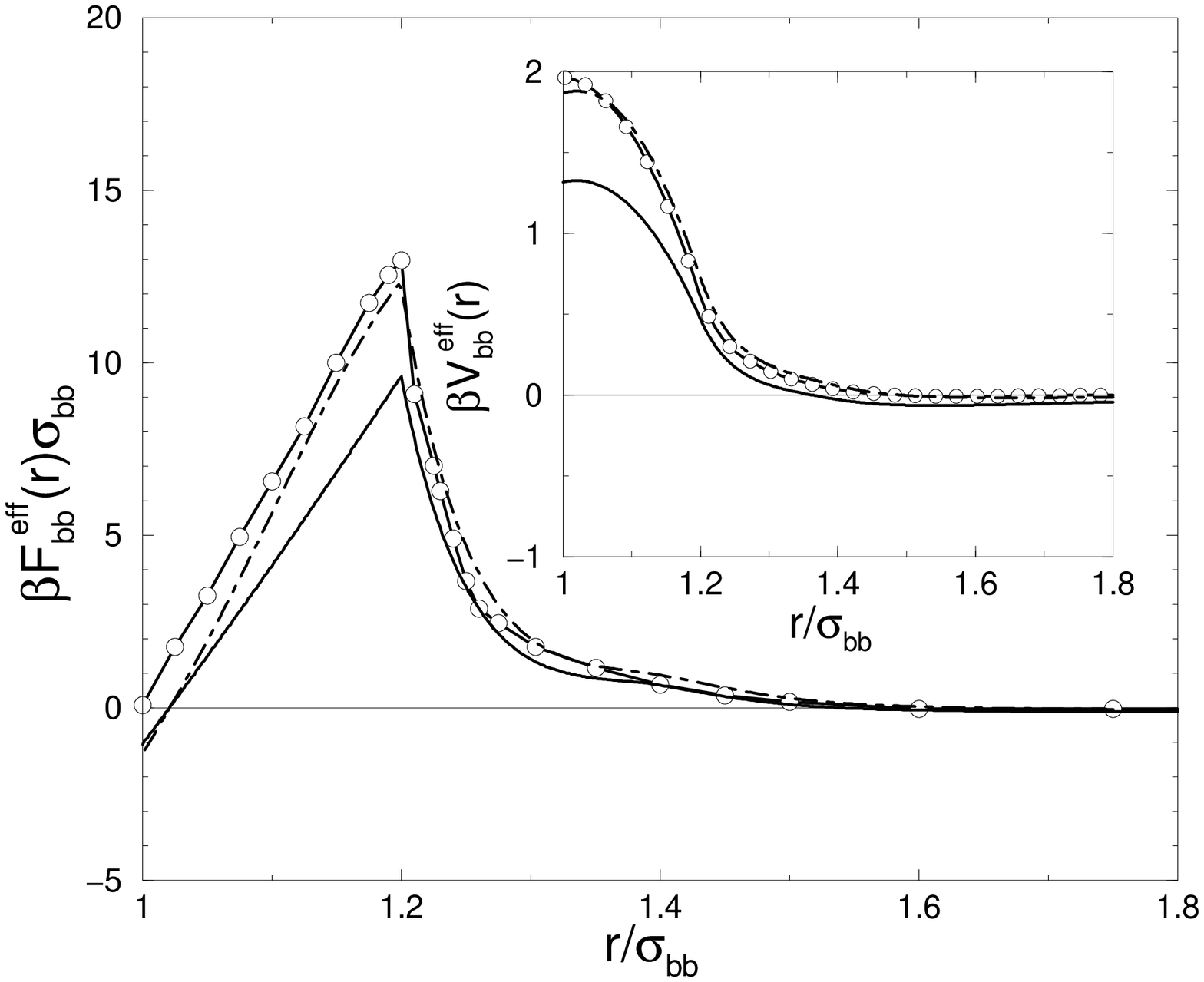,width=8cm} 
\caption{\label{fig:force7}  Same as Fig.~\ref{fig:force1}, but for run7: 
$\epsilon_{bs}<0$, $\epsilon_{ss}=0$.
}
\end{center}
\end{figure}

\begin{figure}
\begin{center}
\epsfig{figure=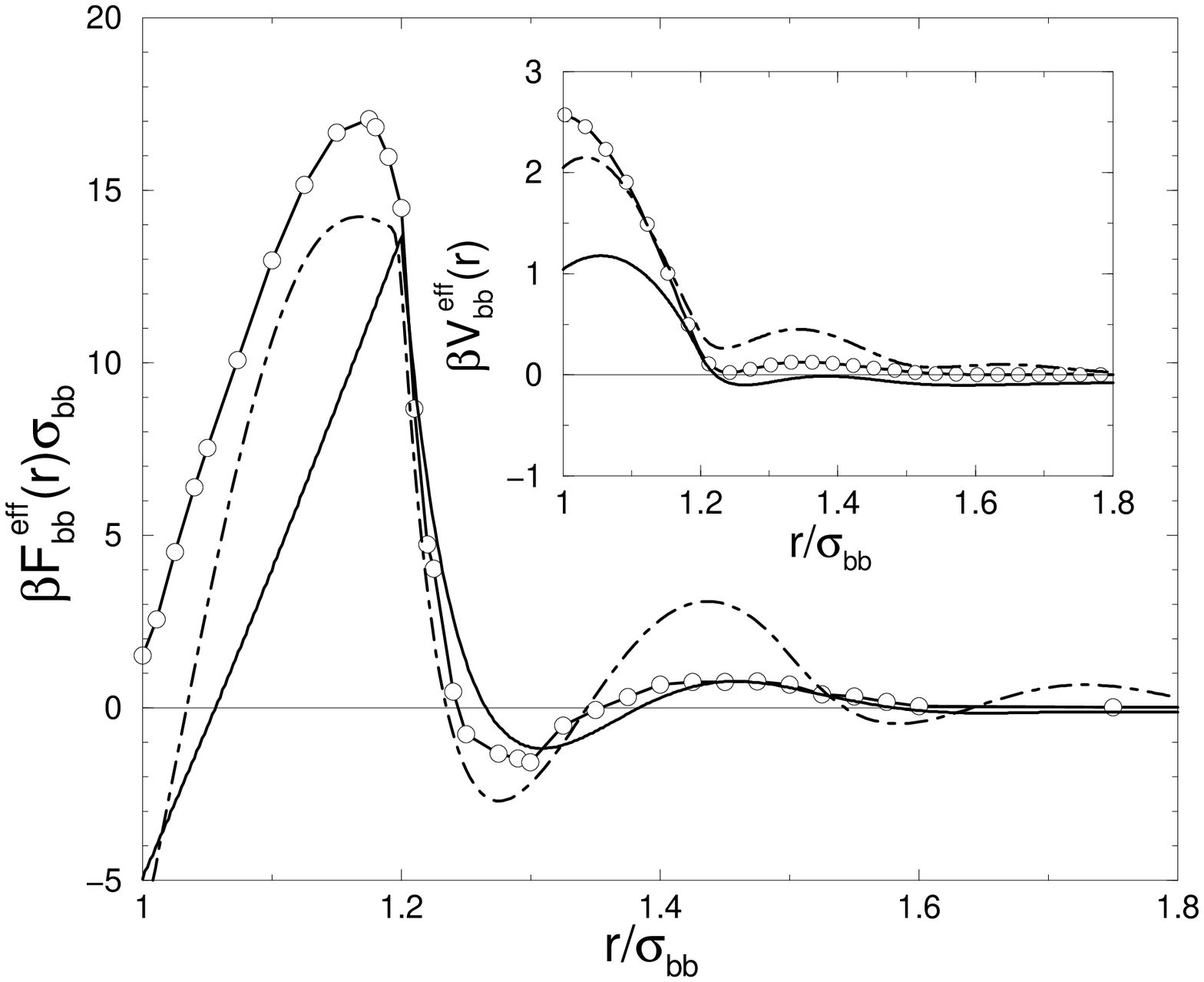,width=8cm} 
\caption{\label{fig:force8}  Same as Fig.~\ref{fig:force1}, but for run8: 
$\epsilon_{bs}<0$, $\epsilon_{ss}>0$.
}
\end{center}
\end{figure}

\begin{figure}
\begin{center}
\epsfig{figure=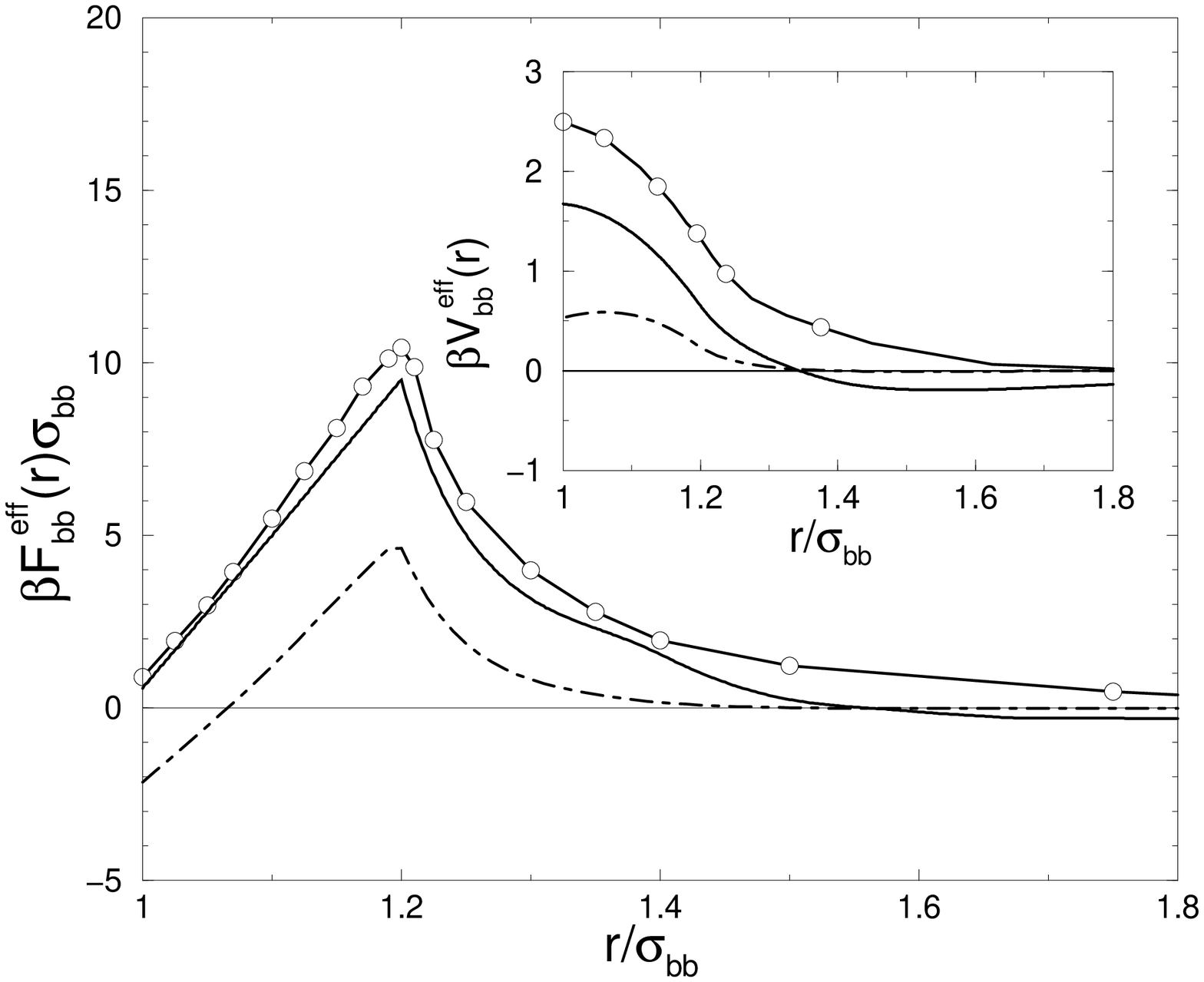,width=8cm} 
\caption{\label{fig:force9} Same as Fig.~\ref{fig:force1}, but for run9: 
$\epsilon_{bs}<0$, $\epsilon_{ss}<0$.
}
\end{center}
\end{figure}

 To calculate the effective forces via Eq.~(\ref{eq3.1}), one needs a
prescription for calculating the one-body density of the small
particles.  In section~\ref{sec:simulation}, we essentially did this
by computer simulations. In this section we approximate the full
one-body density by a superposition of the one-body density
$\rho_s(r)$ around an isolated single sphere\cite{Atta89}:
\begin{equation}\label{eq3.2}
\rho_s^{(1)}({\bf r}';{\bf 0}, {\bf r}) =  \rho_s( r') \rho_s(|{\bf r
- r}'|)/\rho_s,
\end{equation}
an approach similar in spirit to the Kirkwood superposition
approximation\cite{Hans86}.  Since all input information comes from
the (radially symmetric) problem of a single sphere, this
superposition approximation greatly simplifies the calculation of the
two-body depletion forces.  The input $\rho_s(r)$ could come from
density functional theory or integral equation theory, as was done
previously by other authors\cite{Bibe96,Amok98}, but here we will use
the $\rho_s(r)$ generated by our simulations and depicted in
Figs.~\ref{fig:hrB0-new}--\ref{fig:hrBattractive-new}.  The results are shown
in Figs.~\ref{fig:force1}--\ref{fig:force9}, where we compare in
detail the radial force $F^{eff}_{bb}(r) = {\bf F}^{eff}_{bb}(r) \cdot
{\bf r}/r$ and the effective potential $V_{bb}^{eff}(r)$ calculated
with the superposition approximation, to the results obtained by
direct simulations.

Figure~\ref{fig:force1} (run 1) shows the HS reference case.  Here the
superposition approximation works quite well. The packing fraction of
the small spheres $\eta_s = 0.1$ is rather low, so we expect that the
small-small correlation effects are not very strong. The total
one-body density is therefore well approximated by Eq.~(\ref{eq3.2}).
As demonstrated by other authors\cite{Bibe96,Amok98} for the pure HS
case, this superposition approximation begins to break down as
$\eta_s$ increases and two-body correlation effects become more
important. For example, they found that for large $\eta_s$, the
superposition approximation leads to an overestimate of the strength
of the attractive force at contact.  This is exactly what is seen in
Fig.~\ref{fig:force2} (run 2), where the effect of an added repulsive
$\phi_{ss}(r)$ can be understood in terms of a larger effective
packing fraction $\eta_s$.  Figure~\ref{fig:force3} (run 3) shows that
for $\epsilon_{ss} < 0$ the superposition approximation shows a similar
error to what was seen for run 2, i.e. the forces and potentials are
too attractive.

In Figs.~\ref{fig:force4}--\ref{fig:force6} (runs 4-6), where the
repulsive $\epsilon_{bs}(r)$ induces much more attractive effective
interactions than for the pure HS case, the superposition
approximation is seen to work quite well for $\epsilon_{ss} =0$ and
$\epsilon_{ss} >0$.  This is most likely because the effective
(depletion) force or potential is dominated by $\phi_{bs}(r)$, which
also induces a lower $\rho_s(r)$ (see
Fig.~\ref{fig:hrBrepulsive-new}), so that small-small correlations
only play a relatively minor role and Eq.~(\ref{eq3.2}) is rather
accurate.  Even for $\epsilon_{ss} < 0$, the results are quite good,
although the superposition approximation tends to predict forces and
potentials that are too attractive, just as was found for
$\phi_{bs}(r)=0$.

In Figs.~\ref{fig:force7}--\ref{fig:force9} (runs 7-9), where the
attractive $\phi_{bs}(r)$ induces an increased local accumulation of
the small spheres near each big sphere, the superposition
approximation performs less well.  This is perhaps not surprising, as
the enhanced density of small particles near each big sphere results
in a more important role for small-small correlations, which are not
well treated by the superposition approximation.

In summary then, for HS or other repulsive small-small interactions,
the superposition approximation works best for low (effective) packing
fractions $\eta_s$, where correlations between the small particles do
not strongly alter the full one-body density $\rho_s^{(1)}({\bf r}';{\bf
0},{\bf r})$ from the superposition of the densities around an
isolated big sphere.  The case of a strongly repulsive $\phi_{bs}(r)$,
which lowers the effective packing fraction near the big spheres, is
particularly well described by the superposition approximation.  In
contrast, the case of an attractive $\phi_{bs}(r)$, which results in
an increased accumulation of small particles around each big sphere,
is not as well described.

\subsection{Density Functional Theory}

In a recent development, density functional theory (DFT) has been used
to derive effective potentials for additive\cite{Gotz99,Roth00a} and
non-additive\cite{Loui01b,Roth01b,Roth01a} HS mixtures.  For the
additive case, quantitative agreement with computer simulations was
achieved\cite{Gotz99,Roth00a}.  Since the calculations for the
non-additive case were shown to be equivalent to the additive
ones\cite{Loui01b,Roth01b}, a similar quantitative accuracy could be
expected there.  In brief, the method is based on the following exact
relationship between the effective potential and the one-body direct
correlation function\cite{Gotz99,Roth00a}:
\begin{equation} \label{eq3.3}
\beta V^{eff}_{bb}({\bf r}) = \lim_{\mu_b\to -\infty}
(c_b^{(1)}(\infty)-c_b^{(1)}({\bf r}))
\end{equation}
where $c_b^{(1)}({\bf r})$ is defined for the case where a big sphere
is fixed at the origin, and exerts a field on the small spheres and on
a big test particle inserted at ${\bf r}$\cite{Roth00a}.  DFT can
therefore provide a route to the effective potentials since
\begin{equation}\label{eq3.4}
c_b^{(1)}({\bf r}) = - \beta \frac{\delta {\cal
F}_{ex}[\rho_b,\rho_s]} {\delta \rho_b({\bf r})}
\end{equation}
where ${\cal F}_{ex}[\rho_b,\rho_s]$ is the excess (over ideal)
intrinsic Helmholtz free energy functional of the
mixture\cite{Evan79}.  Given some mixture functional, one can obtain
the effective potentials and forces from a radially symmetric
calculation of $c_b^{(1)}({\bf r})$ in the limit of vanishing density
of the big spheres.  This is much simpler than say trying to use DFT
to directly calculate the cylindrically symmetric one-body density
$\rho_s^{(1)}({\bf r}';{\bf 0}, {\bf r})$, for use in
Eq.~(\ref{eq3.1})\cite{vonG99}.  All one now needs is some
prescription for the mixture functional.  Here we use Rosenfeld's very
successful fundamental measure theory\cite{Rose89}, in its form valid
for HS mixtures.  As found previously\cite{Roth00a}, the DFT and
simulation results for effective interactions of the pure HS case
depicted in Fig.~\ref{fig:force1} (run 1) agree very well.

This DFT method can easily be extended to an arbitrary $\phi_{bs}(r)$,
since this simply corresponds to an additional external field in
Eq.~(\ref{eq3.3}).  Previous DFT calculations\cite{Loui01b} found good
agreement with earlier simulations with an attractive
$\phi_{bs}(r)$\cite{Malh99}.  Here we also find very good agreement
with the runs $4$, and $7$, (Figs.~\ref{fig:force4}
and~\ref{fig:force7}) which correspond to a finite $\phi_{bs}(r)$ but
no additional $\phi_{ss}(r)$.

Since there is at present no successful two-component DFT for mixtures
where the small-particles are not HS like, some approximations must be
made.  We chose to map the $\phi_{ss}(r)$ onto effective HS diameters
$\sigma_{ij}^{eff}$ using the Barker-Henderson approach\cite{Bark67}:
\begin{equation}\label{eq3.5}
\sigma_{ij}^{eff} = \sigma_{ij} + \int_{\sigma_{ij}}^\infty (1-\exp[-\beta \phi_{ij}(r)]) dr,
\end{equation} where $\sigma_{ij}$ is the bare HS diameter.
We then calculated effective pair potentials and effective pair forces
with our full DFT approach, including an explicit $\phi_{bs}(r)$ when
needed, but with the small-spheres mapped onto the effective diameters
$\sigma_{ss}^{eff} = 1.447 \sigma_{ss}$ and $\sigma_{ss}^{eff} = 0.658
\sigma_{ss} $ for the repulsive and attractive $\phi_{ss}(r)$
respectively.  Similarly the effective packing fraction becomes
$\eta_{ss}^{eff} = 0.303$ for the repulsive and $\eta_{ss}^{eff} =
0.0290$ for the attractive interaction. 

Firstly, for runs $2$, $5$, and $8$ (Figs.~\ref{fig:force2},
~\ref{fig:force5}, and~\ref{fig:force8}, respectively), which all
correspond to a repulsive $\phi_{ss}(r)$, we find good agreement for
no additional $\phi_{bs}(r)$ (see Fig.~\ref{fig:force2}), but less good
agreement for a repulsive or attractive $\phi_{bs}(r)$ (see
Fig.~\ref{fig:force5} and Fig.~\ref{fig:force8} respectively).

Because of the strength of the small-small attraction, the mapping
results in a very low effective packing fraction of the small
particles.  Overall, the DFT underestimates the effective forces for
$\epsilon_{bs} = 0$ and $\epsilon_{bs} > 0$, as can be seen in
Figs.~\ref{fig:force3} and~\ref{fig:force6}.  It performs rather
poorly for run 9, where the attractive $\phi_{bs}(r)$ results in a
non-linear enhancement of the small-particle density profile, an
effect not taken into account with our HS mapping.  This suggests that
a different two-component DFT, which explicitly takes into account the
small-small attraction $\phi_{ss}(r)$ needs to be developed before we
can use this route to derive accurate effective pair forces and
potentials.

\subsection{Mapping to non-additive HS system}
In a previous paper\cite{Loui01b} two of us proposed that the effects
of big-small and small-small interactions on the $V_{bb}^{eff}(r)$
could be understood by mapping onto those of non-additive HS systems.
These have the advantage that they can be determined by an exact
mapping onto the depletion potentials of additive HS
mixtures\cite{Loui01b,Roth01b}, which, in turn, are well understood
and for which a good parameterization exists\cite{Roth00a}.  Even a
small non-additivity was shown to have a large effect on the depletion
potentials.

By mapping the big-small and small-small interactions onto effective
HS diameters through Eq.~(\ref{eq3.5}) one can  define the
non-additivity in terms of the parameter $\Delta$:
\begin{equation}\label{eq3.10}
\sigma_{bs}^{eff} = \frac12 \left( \sigma_{bb} +
\sigma_{ss}^{eff}\right) \left( 1 + \Delta\right).
\end{equation}Four different ways of 
adding interactions to introduce non-additivity were
studied\cite{Loui01b}:

\noindent{\bf(i)}\,\,\, repulsive $\,\,\beta \phi_{ss}(r)$: $\sigma_{bs}^{eff}=\sigma_{bs}$; $\sigma_{ss}^{eff}> \sigma_{ss}$;
$\Delta < 0$

\noindent{\bf(ii)} attractive $\beta \phi_{ss}(r)$: $\sigma_{bs}^{eff}=\sigma_{bs}$; $\sigma_{ss}^{eff}<\sigma_{ss}$;
$\Delta > 0$

\noindent{\bf(iii)} repulsive  $\beta \phi_{bs}(r)$: $\sigma_{bs}^{eff}>\sigma_{bs}$; $\sigma_{ss}^{eff}=\sigma_{ss}$; $\Delta > 0$

\noindent{\bf(iv)} \hspace*{-6pt} attractive $\beta \phi_{bs}(r)$:
$\sigma_{bs}^{eff}<\sigma_{bs}$; $\sigma_{ss}^{eff}=\sigma_{ss}$;
$\Delta < 0$.

\noindent For details of how each of the different cases of
non-additivity affect the depletion potentials, we refer to
refs.\cite{Loui01b,Roth01b,Roth01a}.

An example where the mapping to non-additivity works well is given in
Fig.~\ref{fig:force2} (run 2), where we mapped the small-small
repulsion onto an effective HS diameter (case {\bf (i)}), so that the
DFT result is really that of a non-additive HS mixture with $\Delta =
-0.069$.  In ref.~\cite{Loui01b} we also found semi-quantitative
agreement with the mapping for weak and short-ranged $\beta
\phi_{bs}(r)$.  In the present case, where $\beta \phi_{bs}(r)$ is
stronger and longer ranged, the agreement is no longer quantitative.
Nevertheless, for runs 1-8 the mapping scheme provides a qualitative
explanation of the trends.  I.e. for a fixed $\phi_{bs}(r)$, adding a
repulsive (case {\bf (i)}) or attractive (case {\bf (ii)})
$\phi_{ss}(r)$ causes $V_{bb}(r)$ to become more repulsive or
attractive, respectively. Similarly, for a $\phi_{ss}(r)$, adding a
repulsive (case {\bf (iii)}) or attractive (case {\bf (iv)})
$\phi_{bs}(r)$ results in a more attractive or repulsive $V_{bb}(r)$,
respectively.  The only case where this scheme breaks down is run 9,
where adding $\phi_{ss}(r) < 0$ for an attractive $\phi_{bs}(r)$
should fall under case {\bf (ii)}.  But instead of inducing more
attraction, the effective big-big potential becomes more repulsive.
Of course this is not surprising, since the DFT results already showed
that a mapping scheme misses the non-linear enhancement of the small
particle density profiles.

In conclusion then, the mapping to non-additivity works best for a
repulsive small-small interaction.  For repulsive and attractive
big-small interactions, the mapping is only qualitative.  The much
better semi-quantitative agreement found in ref.~\cite{Loui01b} can be
traced to the much weaker effect of the $\phi_{bs}(r)$ used there.
Just as was seen for the direct DFT methods, it is the case of
small-small attraction combined with big-small attraction which seems
most difficult to capture within our mapping scheme.

\section{conclusions}

In conclusion, we have shown how the basic interactions in a colloidal
mixture control the resulting effective interactions between the big
particles.  This knowledge may be exploited to stabilize colloidal
particles against coagulation and to tailor the colloidal phase
diagrams.

  Adding a repulsive $\phi_{bs}(r)$ results in a strongly enhanced
attraction through the standard depletion mechanism.  We also found at
least two ways to obtain significantly more repulsive effective
interactions caused by accumulation of the small particles
(accumulation repulsion): {\bf (1)} Adding repulsions between the
small particles results in an enhanced accumulation near the surface
of the large particles; when two large particles approach each other,
this results in an effective repulsion between them.  {\bf (2)} Adding
an attraction between the large and the small particles also results
in an enhanced accumulation near the surface of the large particles
and therefore in repulsive effective interactions.  Furthermore, we
found that for an attractive big-small interaction, adding small-small
attractions resulted in even more effective repulsion.  This
``repulsion through attraction'' effect is caused by a coupling
between $\phi_{ss}(r)$ and $\phi_{bs}(r)$.

 These predictions could in principle be verified in experiments that
directly measure the effective forces of colloidal suspensions, such
as optical tweezers\cite{Croc99,Verm98} or Total Internal Reflection
Microscopy (TIRM)\cite{Rudh98,Bech99,Prie99}.  They could also be
verified indirectly through measurements of phase-behavior and
coagulation.  Measurements of the second osmotic virial coefficient
might also be very sensitive probes of the effective
interactions\cite{Roth01b}.  Systems where these interactions could be
tuned include for example ternary suspensions\cite{Koni01} where the
small-small attraction is generated by depletion attraction of an even
smaller third colloidal component or highly salted charged suspensions
where the van-der-Waals attraction dominates.

In some interesting recent experiments\cite{Tohv01}, a colloidal
suspension of neutral big particles was stabilized by the addition of
highly charged small nanoparticles.  The proposed mechanism was termed
``nanoparticle halos'', and is very similar to the mechanism we
observe, for example, in our run $2$, where a repulsive $\phi_{ss}(r)$
of the screened Coulomb (Yukawa) form was added to the smaller
particles, resulting in an increased accumulation of small particles
near each big one, and an effective repulsion between the big
particles.  The effects in the experiments may also be enhanced by a
small attractive $\phi_{bs}(r)$ (like our run $8$). We are currently
actively pursuing a more detailed comparison with these experiments.

Another possible application of this work is to supercritical
solvents\cite{Kira94}, which have important applications in industrial
processes.  The question of how the effective force on the big
particles depends on the interactions with a low density solvent is
encountered there as well\cite{Kuwa00,Egor01}, sometimes for similar
size ratios.

More generally to make useful predictions for practical applications,
one needs to more thoroughly explore the rather large parameter space,
which includes the $\epsilon_{ij}$, the $\kappa_{ij}$, $\eta_s$ and
the ratio $\sigma_{ss}/\sigma_{bb}$.  This would be very tedious with
simulations -- trustworthy theoretical techniques would be more
practical.  We attempted a number of theoretical descriptions of the
effective interactions.  The superposition approximation, which has
the advantage of only needing one-body input, works best for a strong
big-small repulsion, but becomes much less accurate when the (local)
small particle packing is high.  When the smaller component is purely
HS like, then our DFT approach is very accurate, just as was found for
additive and non-additive HS
mixtures\cite{Roth00a,Loui01b,Roth01b,Roth01a}.  When a repulsive
interaction is added between the small particles, DFT does not perform
quite as well, and when an attraction is added between the small
particles, the differences are even more important.  By mapping the
small-small and big-small interactions onto effective HS diameters, we
can map onto an effective non-additive HS model. This explains the
qualitative trends for most of our parameter combinations.  It is
quantitative if only a small-small repulsion is added, but breaks down
when both a small-small and a big-small interaction couple together to
induce an enhanced small particle density.

All three tested theoretical methods perform less well for added
small-small attractions.  Constructing a reliable theory to treat this
very interesting case is therefore a challenging problem.  One
possible way to extend the DFT calculations to an attractive
small-small interaction would be to add a mean-field attractive
term\cite{Evan79}.  For example, mean-field like functionals based on
thermodynamic perturbation theory around a HS reference system have
been successfully applied to the density profiles and phase-behavior
in systems with attractive potentials \cite{Curt86,Ohne94}.

 New physics is expected when wetting or drying phenomena control the
density of the small particles between the big ones. This is relevant
if the bulk fluid of the small particles is close to liquid-gas phase
coexistence.  A wetting transition is expected to have a profound
impact on the effective interaction as well, one important effect is a
liquid ``bridge'' of small particles between the big one which has
been recently studied in more detail \cite{Biek00}.

We thank Jean-Pierre Hansen, Geoff Maitland, Edo Boek, and Bob Evans
for illuminating discussions.  AAL acknowledges support from the Isaac
Newton Trust, Cambridge, HL acknowledges support from Schlumberger
Cambridge Research, through the Schlumberger visiting professor
scheme.

\begin{table}
\caption{\label{tableI} Parameter combinations for simulation runs.
The interactions are given by Eqs.~(\protect\ref{eq2.1})
and~(\protect\ref{eq2.2}), and only the $\epsilon_{bs}$ and
$\epsilon_{ss}$ are changed; the other parameters are kept fixed.  }
\begin{tabular}[t]{lcc}
 run & $\beta \epsilon_{bs}$ & $\beta \epsilon_{ss}$  \\ 
\hline
\hline
1 & 0  & 0 \\ 
2 & 0  & 2.99 \\ 
3 & 0  & -0.996 \\ 
\hline
4 & 0.82  & 0 \\ 
5 & 0.82  & 2.99 \\ 
6 & 0.82 & -0.996 \\
\hline
7 & -0.82  & 0 \\ 
8 & -0.82  & 2.99 \\ 
9 & -0.82  & -0.996 \\ 
\end{tabular}
\end{table}

i

\end{multicols}


\begin{thebibliography}{99}
\bibitem{nonadd}  In this hard-particle context non-additive means
that the cross-diameter $\sigma_{bs} \neq
\frac12(\sigma_{ss}+\sigma_{bs})$, where the $\sigma_{ij}$ are the
HS diameters of the particles.
\end{thebibliography}
\end{document}